\documentclass[twocolumn, twocolappendix]{aastex631}

\let\tablenum\relax

\usepackage{amsmath}
\usepackage{graphicx}
\usepackage{orcidlink}
\usepackage{txfonts}
\usepackage{hyperref}
\usepackage{booktabs} 
\usepackage{tabularx}
\usepackage{siunitx}
\usepackage{csquotes}
\usepackage{subfiles}
\usepackage{wasysym} 
\usepackage{enumitem}
\usepackage{makecell}
\usepackage{threeparttable}

\usepackage[normalem]{ulem}   
\usepackage{xcolor}           
\usepackage{amsmath,amssymb}  

\newcommand{\e}{\text{e}}
\newcommand{\der}{\text{d}}

\newcommand{\boldmat}[1]{\boldvec{\mathrm{#1}}}
\newcommand{\boldvec}[1]{\boldsymbol{#1}}

\begin{document}


\title{Robust Data Interpretation for Perturbed Nulling Interferometers via Proper Handling of Correlated Errors}

\author[0009-0000-7417-3201]{Philipp A. Huber}
\affiliation{ETH Zurich, Institute for Particle Physics and Astrophysics, Wolfgang-Pauli-Strasse 27, 8093 Zurich, Switzerland}
\affiliation{National Center of Competence in Research PlanetS (\url{www.nccr-planets.ch})}

\author[0000-0002-5476-2663]{Felix A. Dannert}
\affiliation{ETH Zurich, Institute for Particle Physics and Astrophysics, Wolfgang-Pauli-Strasse 27, 8093 Zurich, Switzerland}
\affiliation{National Center of Competence in Research PlanetS (\url{www.nccr-planets.ch})}

\author[0000-0002-2215-9413]{Romain Laugier}
\affiliation{Institute of Astronomy, KU Leuven, Celestijnenlaan 200D, 3001, Leuven, Belgium}

\author[0009-0003-7304-7512]{Taro Matsuo}
\affiliation{Department of Particle and Astrophysics, Graduate School of Science, Nagoya University Furocho, Chikusa-ku, Aichi, Nagoya, 466-8601, Japan}

\author[0009-0002-6295-4558]{Loes W. Rutten}
\affiliation{Institute of Astronomy, KU Leuven, Celestijnenlaan 200D, 3001, Leuven, Belgium}

\author[0000-0001-9250-1547]{Adrian M. Glauser}
\affiliation{ETH Zurich, Institute for Particle Physics and Astrophysics, Wolfgang-Pauli-Strasse 27, 8093 Zurich, Switzerland}
\affiliation{National Center of Competence in Research PlanetS (\url{www.nccr-planets.ch})}

\author[0000-0003-3829-7412]{Sascha P. Quanz}
\affiliation{ETH Zurich, Institute for Particle Physics and Astrophysics, Wolfgang-Pauli-Strasse 27, 8093 Zurich, Switzerland}
\affiliation{National Center of Competence in Research PlanetS (\url{www.nccr-planets.ch})}
\affiliation{ETH Zurich, Department of Earth Sciences, Sonneggstrasse 5, 8092 Zurich, Switzerland}

\collaboration{20}{(LIFE Collaboration)}



\begin{abstract}
    The detection and atmospheric characterization of potentially habitable, temperate terrestrial exoplanets using a space-based mid-infrared nulling interferometer is a major goal of contemporary astrophysics. A central part of the analysis of such an instrument are correlated errors arising from perturbations in the system. While previous studies have often treated their effects in a limited manner, we aim to treat them comprehensively here and argue that data whitening based on the covariance of these errors is a suitable method to mitigate their impact. We present a framework that quantitatively connects instrumental perturbations to performance metrics and develop two computational tools to support our analysis: \textsc{PHRINGE}, for the generation of synthetic nulling data, and \textsc{LIFEsimMC}, a new Monte Carlo-based end-to-end simulator for the Large Interferometer For Exoplanets (LIFE). Applying our framework to a reference observation of an Earth twin orbiting a Sun twin at \SI{10}{pc}, we find that whitening is not only essential for a correct interpretation of the detection metric used in hypothesis testing, but also improves the estimates of the planetary properties. Moreover, our approach enables an estimation of the spectral covariance of the extracted planetary spectra, providing valuable additional input for future atmospheric retrievals. We therefore recommend incorporating the framework into performance assessments and requirement derivations for future nulling interferometers.

\end{abstract}

\keywords{Astronomical simulations (1857) --- Monte Carlo methods (2238) --- Astronomy software (1855) --- Exoplanet astronomy (486) --- Direct detection interferometry (386) --- Space telescopes (1547)}


%
%

    \section{Introduction}\label{sec:intro}
    A major goal of contemporary astrophysics is the detection and atmospheric characterization of temperate terrestrial exoplanets orbiting in habitable zones of nearby stars through their thermal emission \citep{cockell_darwin_2009, quanz_atmospheric_2022, kammerer_large_2022}. Ultimately, this would pave the way towards an understanding of the habitability and diversity of these worlds, enabling the search for atmospheric signatures indicative of biological activity. Space-based nulling interferometers with their unique combination of high angular resolution, high achievable contrasts, and high mid-infrared (MIR) sensitivity are considered a promising option to tackle this challenging task \citep{bracewell_detecting_1978, bracewell_searching_1979, Angel_1997, mennesson_array_1997, mennesson_direct_2005}. Building on the heritage of NASA's TPF-I \citep{lawson_terrestrial_2007} and ESA's Darwin \citep{cockell_darwin_2009} mission concepts, the Large Interferometer For Exoplanets (LIFE) initiative is reevaluating the potential of a MIR nulling interferometer mission leveraging the latest scientific and technological advances. A particular effort of the LIFE community is aimed at deriving detailed technical specifications of the instrument \citep[for a recent set of specifications see][]{glauser_large_2024}.
    End-to-end simulations play a pivotal role in this process \citep{dannert_large_2022}. They link planetary models to instrument models, generate synthetic data of simulated observations, and estimate planet properties from those data through a process called signal extraction. These estimated properties include the planetary radii, effective temperatures, and spectra. The spectra are typically used as input for atmospheric retrieval tools, allowing an estimate of atmospheric structures and compositions \citep[e.g.][]{alei_large_2022, Konrad_2024}. Using this forward modeling approach, we can assess whether a given set of technical specifications meets the scientific objectives of the mission. A key aspect of these simulations is the noise sources that are considered. In this work, we focus on so-called \textit{instability noise}, which originates from the perturbed nature of the nulling instrument \citep{lay_systematic_2004, lay_removing_2006}. In the following, we will also refer to it as \textit{instrumental errors}. Most previous LIFE performance studies assumed that the instrument is built such that it is photon-noise-limited and, hence, did not model instrumental noise explicitly \citep{quanz_large_2022}. Instead, they implicitly accounted for it by assuming an ideal instrument but requiring a signal-to-noise ratio of $S/N\geqslant 7$ instead of the typically used $S/N \geqslant 5$ to claim a exoplanet detection. As we will see, in doing so, they may have imposed overly strict instrumental requirements.\par
    Explicitly modeling instrumental errors requires not only more comprehensive instrument models, but, due to the correlated nature of such errors, also suitable data processing techniques \citep{huber_analytical_2024}. \cite{serabyn_nulling_2000} and \cite{lay_systematic_2004} were the first to tackle this challenge with their efforts to model instrumental errors and connect them to stability requirements for the instrument. \cite{lay_systematic_2004} developed a complex mathematical framework that allowed an analytical calculation of the variance of instrumental errors resulting from perturbations in beam amplitude, phase, and polarization angle. Part of the complexity of the framework is owed to the explicit accounting for temporal correlations in the errors. Emphasis is put on the issue that these correlations can indeed mimic a planetary signal in the data, leading to wrongly claimed detections. In this context, \cite{dannert_consequences_2025} reassess the typical assumption of Gaussian distributed instrumental errors used in hypothesis tests to inform such detection claims. They show that this assumption is invalid and derive an analytical expression for a hypothesis test that accounts for the true distribution of the errors.\par
    While this work is essential for a correct interpretation of the measurements, it only addresses part of the issue of correlated errors: both, \cite{lay_systematic_2004} and \cite{dannert_consequences_2025}, omit a treatment of spectral correlations. First efforts to explicitly model and mitigate the impact of said spectral correlations was done by \cite{lay_removing_2006}. The author presented an ad hoc approach of first fitting and then subtracting the errors from the data. The fidelity of this approach was increased by \cite{matsuo_large_2023} in the context their proposed signal extraction algorithm, called phase-space synthesis decomposition, entailing an extrapolation (from short towards longer wavelengths) and successive subtraction of the errors. Continuing these developments, a more rigorous approach was followed by \cite{laugier_asgardnott_2023} for the ground-based nulling instrument Asgard/NOTT. They applied a \textit{whitening transformation} proposed by \cite{ceau_kernel_phase_2019} in a slightly different context to mitigate the impact of the correlations on the detection limits of their instrument. This whitening transformation is also the method that is used in this work.\par
    Modeling and correctly handling correlated instrumental errors is essential for a robust interpretation of the generated data, which in turn is an essential prerequisite for a robust derivation of technical specifications. In this study, we thus aim to establish a means to treat spectral correlations by arguing that \textit{(1) data whitening is a suitable method to mitigate their effects and (2) applying this method shows prospects for relaxing technical requirements and providing additional information that can potentially aid subsequent atmospheric retrievals from the estimated spectra}. We first develop a new state-of-the-art synthetic data generation tool, called \textsc{PHRINGE}, tailored to the use case of perturbed space-based nulling interferometers. An overview of the instrument model implemented within \textsc{PHRINGE} and a description of the correlations in the instrumental errors it produces is given in Section \ref{sec:data}. An explanation of the data whitening transformation, hypothesis testing and signal extraction procedure is outlined in Section \ref{sec:data_processing_signal_extraction}. There, we also describe our second tool, \textsc{LIFEsimMC}, a new end-to-end simulator for LIFE that incorporates these data processing techniques. In Section \ref{sec:results}, we apply our framework to the reference case of an Earth twin orbiting a Sun twin at \SI{10}{pc}. We then discuss our findings and limitations of the framework in Section \ref{sec:discussion} and finally conclude our work in Section \ref{sec:conclusion}.

    \section{Data from perturbed instruments}\label{sec:data}
    
    \subsection{Data Model}\label{subsec:data_model}
\begin{figure}
        \centering
        \includegraphics[width=0.8\linewidth]{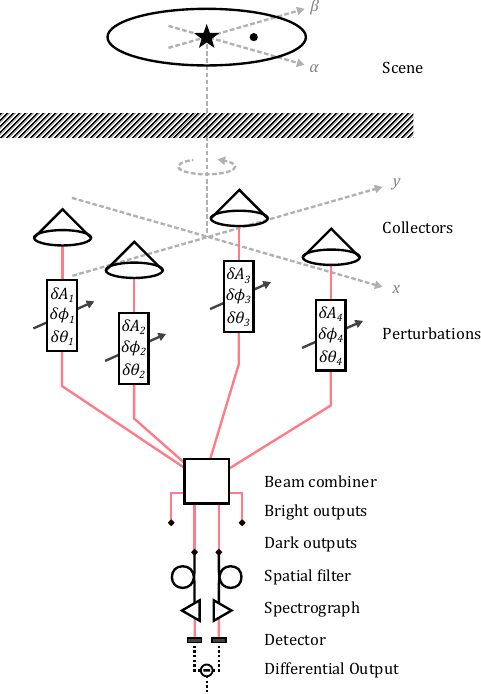}
        \caption{Definition of the instrumental setup. The observed astrophysical scene consists of the star, the planet, the exozodiacal dust (exozodi; ellipse) and the local zodiacal dust (local zodi; shaded area). The four collectors are shown in the X-array configuration, rotating around the line-of-sight, which is centered on the star to \textit{null} its light. The shorter baseline between two collectors is called nulling baseline, while the longer one is called imaging baseline. Amplitude, phase, and polarization angle perturbations are introduced in the input beams. The beams are combined in the beam combiner to form different outputs. For a double Bracewell nuller, four inputs are combined into two bright (left and right) and two dark (bottom) outputs. The dark outputs are spatially filtered using a single mode fiber (SMF), fed into a spectrograph and measured on the detectors. The measured signals are then combined in post-processing to form the so-called differential output. This figure is adapted from \cite{lay_systematic_2004}.}
        \label{fig:setup1}
    \end{figure}

    The definition of the setup used in the following derivations is given in Figure \ref{fig:setup1}. 
    The (photoelectron) counts that we register at the $j$th (bright or dark) output of the beam combiner at time $t$ and wavelength $\lambda$ can be described as
    \begin{equation}\label{eq:photon_counts}
        N_j(t, \lambda)=\eta_\text{QE} w(\lambda) t_\text{DIT}\iint R_j(t, \lambda, \boldvec{\alpha})I(\lambda, \boldvec{\alpha}) \der \boldvec{\alpha},
    \end{equation}
    where $\eta_\text{QE}$ is the quantum efficiency of the detector in electrons per photon, $w$ is the width of the wavelength bin with center wavelength $\lambda$ in \si{\mu m}, $t_\text{DIT}$ is the detector integration time in \si{s}, $R_j$ is the $j$th instrument response in \si{m^2}, $\boldvec{\alpha}=(\alpha, \beta)$ are the sky coordinates in \si{rad}, and $I$ is the sky brightness distribution of the astrophysical scene in \si{ph.s^{-1}.m^{-2}.\mu m^{-1}.sr^{-1}}. Specifically, Equation (\ref{eq:photon_counts}) describes the counts collected in the time interval $[t, t+t_\text{DIT}]$ and wavelength interval $[\lambda-w(\lambda)/2, \lambda+w(\lambda)/2]$. This assumes that both $t_\text{DIT}$ and $w(\lambda)$ are sufficiently small, such that the signal is constant within those intervals. The units of \si{m^2} of the instrument response, $R_j$, suggest that it can be thought of as the \enquote{effective} collecting area of the interferometer. In more technical terms, it corresponds to the amplitude response of the instrument at output $j$ to a point source with unit spectral energy distribution (SED) that is located at $\boldvec{\alpha}$. As illustrated in Figure \ref{fig:setup1}, output $j\in [1, n_\text{out}]$ is constructed by mixing different interferometric inputs, which we denote by the indices $k\in [1, n_\text{in}]$.
    The mixing operation is described by the beam combination transfer matrix, $\boldmat{M}$. For the double Bracewell nuller considered in this study, $\boldmat{M}$ is defined as
    \begin{equation}\label{eq:beam_combiner}
        \boldmat{M}=\frac{1}{2}\begin{bmatrix}
            0 & 0 & \sqrt{2} & \sqrt{2}\\
            \sqrt{2} & \sqrt{2} & 0 & 0\\
            1 & -1 & -\exp\left(i\pi/2\right) & \exp\left(i\pi/2\right)\\
            1 & -1 & \exp\left(i\pi/2\right) & -\exp\left(i\pi/2\right)
        \end{bmatrix},
    \end{equation}
    as described in \cite{dannert_large_2022}. Note that this is a semi-unitary \citep{guyon_optimal_2013} matrix of shape $n_\text{out}\times n_\text{in}$. The light beams at the inputs of the interferometer are described by their complex amplitudes. For an ideal instrument with unperturbed input beams this is
    \begin{equation}\label{eq:complex_amplitude_ideal}
        \boldvec{E}_{k, \text{ideal}}(t, \lambda, \boldvec{\alpha})=A_k\exp\left(i\frac{2\pi}{\lambda}\boldvec{\alpha}\boldvec{x}_k(t)\right)\Bigg[\begin{matrix}
                \cos\theta_k\\
                \sin\theta_k
            \end{matrix}\Bigg],
    \end{equation}
    where $A_k$ is the amplitude in \si{m^2}, $\boldvec{x}_k(t)=(x_k(t), y_k(t))$ the position of the $k$th collector in the local coordinate frame in \si{m}, and $\theta_k$ the polarization rotation angle in \si{rad}. The vector part of this expression accounts for the pointing of the electric field \citep{lay_systematic_2004}. The amplitude $A_k$ can be written as $A_k=r_k\sqrt{\pi\eta_\text{tot}}$, where $r_k$ are the collector radii in $\si{m}$, $\pi$ accounts for the circular shape of the collectors and $\eta_\text{tot}$ is the unitless total throughput of the system.\par
    To model the response of a perturbed instrument, we introduce terms for perturbations in amplitude, phase, and polarization into Equation (\ref{eq:complex_amplitude_ideal}) and get
        \begin{equation}\label{eq:complex_amplitude}
        \begin{aligned}
            \boldvec{E}_k(t, \lambda, \boldvec{\alpha})&=A_k\left(1+\delta A_k(t)\right)\exp\left(i\left(\frac{2\pi}{\lambda}\boldvec{\alpha}\boldvec{x}_k(t)+\delta\phi_k(t, \lambda)\right)\right)\\
            &\cdot\Bigg[\begin{matrix}
                \cos\left(\theta_k+\delta\theta_k(t)\right)\\
                \sin\left(\theta_k+\delta\theta_k(t)\right)
            \end{matrix}\Bigg].
        \end{aligned}
    \end{equation}
    Here, $\delta A_k$ is the unitless amplitude perturbation, $\delta\phi_k$ is the phase perturbation in \si{rad}, and $\delta\theta_k$ is the polarization angle perturbation in \si{rad}. The intensity response of a perturbed instrument then reads
    \begin{equation}\label{eq:instrument_response}
        \begin{aligned}
            R_j(t, \lambda, \boldvec{\alpha})
            =\Bigg|\sum_{k=1}^{n_\text{in}} M_{jk}E_{k, x}(t, \lambda, \boldvec{\alpha})\Bigg|^2+\Bigg|\sum_{k=1}^{n_\text{in}} M_{jk}E_{k, y}(t, \lambda, \boldvec{\alpha})\Bigg|^2,
        \end{aligned}
    \end{equation}
    where $M_{jk}$ are the elements of $\boldmat{M}$, and $E_{k, x}$ and $E_{k, y}$ are the $x$- and $y$-components of $\boldvec{E}_k$, respectively.
        \begin{figure}
        \centering
        \includegraphics[width=\linewidth]{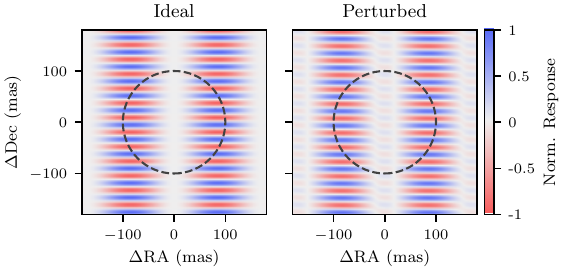}
        \caption{Differential intensity response, $R_4-R_3$, of a double Bracewell nuller in the case of an ideal (left) and a perturbed (right) instrument at $\SI{10}{\mu m}$ at a given instant in time. The impact of the perturbations is highly exaggerated ($\text{RMS}(\delta A_k)=\SI{10}{\%}$, $\text{RMS}(\delta \phi_k)\approx\SI{0.28}{rad}$, $\text{RMS}(\delta \theta_k)=\SI{0.4}{rad}$) here for illustrative purposes. The dashed line corresponds to the path that a planet traces out as the interferometer array rotates around its line-of-sight to the star.}
        \label{fig:intensity_responses}
    \end{figure}
    Figure \ref{fig:intensity_responses} illustrates how amplitude, phase, and polarization angle perturbations lead to perturbations in the instrument response. It shows the differential response, $R_4-R_3$, of an ideal (left) and a perturbed (right) double Bracewell nuller at a given time and wavelength. The intensity response scales with wavelength, resulting in a spacing of $\lambda/2b$ between successive maxima of the fringes, where $b$ is the imaging or nulling baseline length, respectively.

    \subsection{Noise Sources}\label{subsec:noise_sources}

    \subsubsection{Fundamental Noise}
    The fundamental noise sources considered here are of purely astrophysical nature. Specifically, we consider contributions from the host star of an observed system, its exozodi, and the local zodi, as shown in Figure \ref{fig:setup1}. We model them as additional photon sources within our astrophysical scene and hence describe them through their sky brightness distributions. The total sky brightness $I$, that enters Equation (\ref{eq:photon_counts}) can thus be written as
    \begin{equation}\label{eq:astro_noise}
        I(\lambda, \boldvec{\alpha})= I_\text{p}(\lambda, \boldvec{\alpha})+I_\star(\lambda, \boldvec{\alpha})+ I_\text{ez}(\lambda, \boldvec{\alpha})+ I_\text{lz}(\lambda, \boldvec{\alpha}),
    \end{equation}
    where $I_\text{p}$ is the sky brightness distribution of the observed planet, $I_\star$ the one of its host star, $I_\text{ez}$ the one of the exozodi and $I_\text{lz}$ the one of the local zodi. Note that we expect contributions from these astrophysical noise sources even in an ideal instrument. Indeed, due to the high resolution of the instrument, the star starts to be resolved, which prevents it from being perfectly \textit{nullled} in the central destructive fringe of the instrument response (see Figure \ref{fig:intensity_responses}). Since the photon emission from the star is a statistical process that follows a Poisson distribution, photon counts in the outputs of the interferometer may differ, which results in a non-zero contribution in the differential output, often referred to as stellar \textit{leakage}. Similarly, the exozodi leaks into the instrument. The local zodi is a diffuse source that emits throughout the field of view of the instrument, also leaking onto the detector.\par
    The situation is different for the planet: Its angular extent is still too small to be resolved and we can model it as a point source with sky brightness distribution
    \begin{equation}\label{eq:planet_sky_brightness}
        I_\text{p}(\lambda, \boldvec{\alpha})=F_\text{p}(\lambda)\delta \left(\boldvec{\alpha} - \boldvec{\alpha}_\text{p}\right),
    \end{equation}
    where $F_\text{p}$ is the SED of the planet, $\delta$ is the Dirac delta function and $\boldvec{\alpha}_\text{p}$ is the position of the planet in the sky. We use the same models as in \cite{dannert_large_2022} for the astrophysical noise terms in Equation (\ref{eq:astro_noise}). For completeness, they are given in Appendix \ref{sec:appendix_astro_noise}.

    \subsubsection{Instrumental Noise}\label{subsubsec:perturbations}
    As described in Equation (\ref{eq:complex_amplitude}), we introduce instrumental errors by adding amplitude, phase, and polarization angle perturbations to the four input beams of the interferometer. This simple approach is justified by the following two assumptions: First, we assume an SMF that acts as a perfect spatial filter in the instrument. This allows us to reduce spatially resolved wavefront errors to an amplitude and phase error. Secondly, we assume that our optical system downstream of the beam combiner entrance is linear. This then allows us to express perturbations downstream of that entrance as perturbations originating upstream of it, i.e., as perturbations in amplitude, phase, and polarization angle of the beams at the instrument inputs. Essentially, this treats everything downstream of the beam combiner entrance as perturbation-free. \cite{dannert_consequences_2025} point out that this assumption is critical, as, in a double Bracewell combiner, perturbations from those downstream optics would affect the input beams in the same manner and thus result in correlated perturbations within them. However, it can be shown that in the double Bracewell combiner, correlations originating downstream of the combiner entrance have no significant effect \citep[][]{dannert_consequences_2025, lay_systematic_2004}.\par

\begin{figure}
        \includegraphics[width=\linewidth]{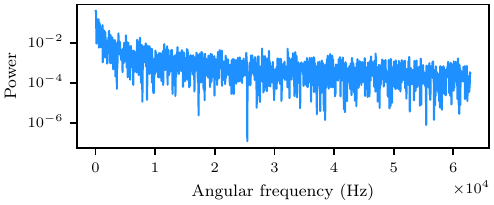}
        \includegraphics[width=\linewidth]{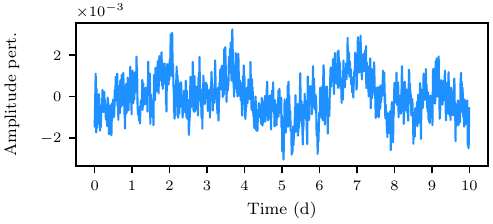}\vspace{0.1cm}
        \caption{\textit{Top:} Underlying pink power spectrum with a $1/\omega$ falloff. \textit{Bottom:} Random time series of an amplitude perturbation with an RMS of $\SI{0.1}{\%}$. }
        \label{fig:psd_time_series}
    \end{figure}

    To create synthetic data for an observation, we generate a random time series for each perturbation term in Equation (\ref{eq:complex_amplitude}). These time series are specified by their root mean square (RMS) and their underlying power spectrum, $S$. In this work, we generate a power spectrum corresponding to pink noise, using
    \begin{equation}\label{eq:psd}
            S(\omega)\propto \frac{1}{\omega},
    \end{equation}
    where $\omega=2\pi f$ is the angular frequency (not to be confused with the rotation frequency of the array). The power spectrum is calculated between a low- and a high-frequency cutoff, $f_\text{low}$ and $f_\text{high}$, respectively. In our simulations, we use $f_\text{low}=1/P_\text{rot}$, where $P_\text{rot}$ is the rotation period of the array, and $f_\text{high}=\SI{10}{kHz}$. The definition of the amplitude perturbation, $\delta A_k$, in Equation (\ref{eq:complex_amplitude}) allows for values in the range $[-1, 1]$. An example of a time series for an amplitude perturbation with an RMS of $\SI{0.1}{\%}$ based on a pink power spectrum is given in Figure \ref{fig:psd_time_series}.
    Note that we model the amplitude perturbations as wavelength-independent here. We define the phase perturbations, $\delta \phi_k$, as
    \begin{equation}\label{eq:phase_pert}
        \delta\phi_k(t, \lambda)=\frac{2\pi}{\lambda} \delta p(t),
    \end{equation}
    where $\delta p$ the piston in units of \si{m}. Finally, the polarization angle perturbations, $\delta \theta_k$, account for mismatches in the linear polarization of the beams and are specified in radians.
    
    \subsection{Differential Signal and Instrumental Errors}\label{subsec:imprint}
    
\begin{figure}
        \includegraphics[width=\linewidth]{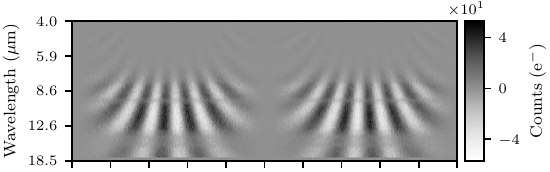}\vspace{-0.1cm}
        \includegraphics[width=\linewidth]{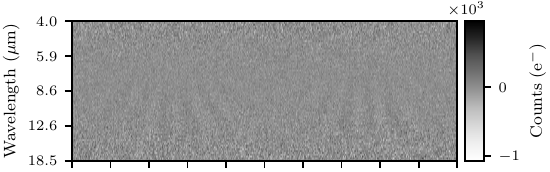}\vspace{-0.1cm}
        \includegraphics[width=\linewidth]{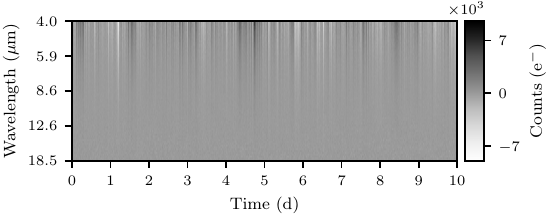}\vspace{-0.1cm}
        \caption{Data obtained from the differential output of a double Bracewell nuller, showing the recorded counts as a function of wavelength and time. \textit{Top}: Signal of an Earth twin at \SI{10}{pc} as recorded by an ideal instrument. \textit{Middle}: Total signal of all astrophysical sources, including the star, the local zodi and the exozodi, as recorded by an ideal instrument. The counts are almost two orders of magnitude higher than for the pure planet signal and enter the data as white noise, predominantly at the low and the high wavelengths. \textit{Bottom}: Total signal of all astrophysical sources as recorded by a perturbed instrument. We have assumed an RMS amplitude perturbation of $\SI{0.1}{\%}$, an RMS phase perturbation of $\SI{1.5}{nm}$ and an RMS polarization angle perturbation of $\SI{0.001}{rad}$. The counts are again almost an order of magnitude higher and introduce strong spectral, but also temporal correlations into the data, as apparent by the vertical stripes.}
        \label{fig:photon_counts_correlations}
    \end{figure}

    As we have anticipated in Section \ref{subsec:data_model}, the science signal of the measurement is obtained by subtracting pairs of signals from two dark outputs,
    \begin{equation}\label{eq:differential_null}
        \Delta N_m(t, \lambda) \equiv N_{k_1}(t, \lambda) - N_{k_2}(t, \lambda), 
    \end{equation}
    where $m\in[1, n_\text{diff}]$ is the index of the differential output and $k_1$, $k_2$ are the indices of the two dark outputs. These differential outputs are more robust against certain instrumental errors, as the dark outputs contain correlated errors that disappear in their difference \citep{martinache_kernel-nulling_2018, laugier_asgardnott_2023}. This is referred to as self-calibration \citep[e.g.][]{hanot_self_2011}. For the double Bracewell nuller used in this study, there is only one differential output ($n_\text{diff}=1$) given by
    \begin{equation}\label{eq:differential_counts_dbw}
        \Delta N_1(t, \lambda) = N_3(t, \lambda) - N_4(t, \lambda).
    \end{equation}
    The top panel of Figure \ref{fig:photon_counts_correlations} illustrates the differential signal of an Earth-like planet at \SI{10}{pc} as recorded by an ideal instrument, i.e., in the absence of instrumental errors.
    The astrophysical and mission parameters used to generate this illustration are given in Table \ref{tab:mission_parameters}.
    \begin{table}[ht!]
        \begin{threeparttable}[b]
        \caption{Astrophysical, instrument, and mission parameters used for the simulations.}
        \label{tab:mission_parameters}
        \begin{tabularx}{\linewidth}{lll}\toprule
            Symbol & Value & Description \\\midrule
            $R_\text{p}$ & \SI{1}{R_\oplus} & Planet radius\\
            $R_\star$ & \SI{1}{R_\odot} & Star radius\\
            $T_\star$ & $\SI{5780}{K}$ & Effective star temperature\\
            $L_\star$ & $\SI{1}{L_\odot}$ & Star luminosity\\
            $d_\star$ & $\SI{10}{pc}$ & Star distance\\
            $\lambda_\text{rel}$\tnote{a} & $\SI{135}{deg}$ & Relative ecliptic longitude\\
            $\beta_\text{lat}$\tnote{a} & $\SI{30}{deg}$ & Ecliptic latitude\\
            $z$ & 3 zodi\tnote{b} & Exozodi level\\\midrule
            $t_\text{int}$ & $\SI{10}{d}$\tnote{c} & Total integration time\\
            $t_\text{DIT}$ & $\SI{600}{s}$ & Detector integration time\\
            $P_\text{rot}$ & $\SI{10}{d}$\tnote{d} & Array rotation period\\
            $b_\text{i}/b_\text{n}$ & 6:1 & Imaging-to-nulling baseline ratio\\
            $b_\text{n}$ & $\SI{9.9}{m}$\tnote{e} & Nulling baseline length\\
            $D$ & $\SI{2}{m}$\tnote{f} & Aperture diameter\\
            $R$ & 50\tnote{g} & Spectral resolving power\\
            $\lambda_\text{min}$ & $\SI{4}{\micro m}$ & Minimum wavelength\\
            $\lambda_\text{max}$ & $\SI{18.5}{\micro m}$& Maximum wavelength\\
            $\eta_\text{tot}$ & $\SI{5}{\%}$\tnote{f} & Total instrument throughput\\
            $\eta_\text{QE}$ &  $\SI{70}{\%}$ & Detector quantum efficiency\\
            \bottomrule
        \end{tabularx}
        \begin{tablenotes}
            \item [a] Only relevant here for local zodi contribution (see Equation (\ref{eq:appendix_local_zodi}))
            \item [b] Level of zodiacal dust emission
            \item [c] The integration time of \SI{10}{d} is chosen arbitrarily, but is of the expected order of magnitude for the given reference observation \citep{konrad_large_2022}.
            \item[d] Corresponding to one array rotation per observation. 
            \item [e] Optimized for maximum sensitivity at the center of the empirical habitable zone of the star at $\lambda = \SI{10}{\mu m}$.
            \item [f] Chosen here for a conservative estimate. Both values are linked to the sensitivity and should thus be considered together.
            \item [g] The spectral resolving power outlined in \cite{glauser_large_2024} is $R=100$, however, we use $R=50$ here for computational reasons.
        \end{tablenotes}
        \end{threeparttable}
    \end{table}
    This signal would also look very similar for a perturbed instrument, as the planetary signal is not strongly influenced by the perturbations. The pattern in the signal is caused by the rotation of the array during the observation (see Figure \ref{fig:intensity_responses}). Certain absorption bands and a generally low blackbody emission at shorter wavelengths lead to a reduction in the signal at specific wavelengths. Figure \ref{fig:flux_and_counts} in Appendix \ref{sec:appendix_flux_and_counts} illustrates this further by showing also the signal for a unit SED and a pure blackbody spectrum. For the signal extraction in Section \ref{sec:data_processing_signal_extraction} it will prove useful to derive an explicit expression for the planetary signal, $\Delta N_{1, \text{p}}$. Using Equations (\ref{eq:planet_sky_brightness}) and (\ref{eq:differential_counts_dbw}), we find
    \begin{align}
            \Delta N_{1, \text{p}}(t, \lambda)&=\eta_\text{QE}w(\lambda)t_\text{DIT}F_\text{p}(\lambda)\left(R_3-R_4\right)\left(t, \lambda, \boldvec{\alpha}_\text{p}\right)\\
            &\equiv F_\text{p}(\lambda)T_\text{p}\left(t, \lambda, \boldvec{\alpha}_\text{p}\right)\label{eq:template},
    \end{align}
    where we have defined the planet model or template function, $T_\text{p}$.\par
    The center panel of Figure \ref{fig:photon_counts_correlations} shows the total signal in the differential output, including contributions from the star, the local zodi, and the exozodi, but still without instrument perturbations. This introduces shot noise in the data that is purely random in nature. \cite{dannert_large_2022} point out that in terms of contributions to shot noise, the exozodi is relatively constant over the full wavelength bandwidth, while the star dominates at the lower wavelengths and the local zodi at the higher ones.
    
\begin{figure}
        \centering
        \includegraphics[width=\linewidth]{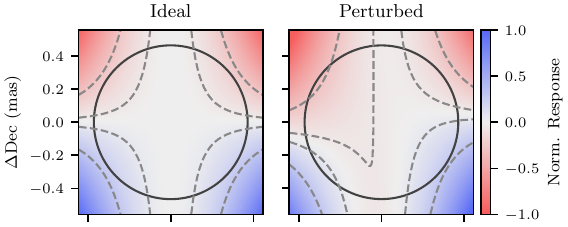}\vspace{0.06cm}
        \includegraphics[width=\linewidth]{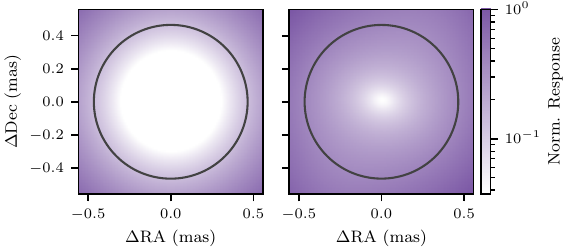}
        \caption{Intensity response within the angular extent of the star (dark gray circle) for an ideal (left) and a perturbed (right) instrument using the values from Table \ref{tab:mission_parameters}, and optimistic perturbation levels of $\text{RMS}(\delta A_k)=\SI{0.1}{\%}$, $\text{RMS}(\delta p_k)\approx\SI{1.5}{nm}$, $\text{RMS}(\delta \theta_k)=\SI{0.001}{rad}$ (see Section \ref{sec:results}). \textit{Top:} Response at a given instance in time. Whereas the null of the ideal instrument is broad and symmetric, the perturbed instrument shows asymmetries, indicated by contours in light gray dashed lines. \textit{Bottom:} RMS of the response over a full rotation of the array. While the ideal response shows a substantial suppression within the extent of the stellar disk, the suppression is much smaller in the perturbed response, explaining the increase in stellar leakage.}
        \label{fig:intensity_response_star}
    \end{figure}

    \paragraph{Increased Stellar Leakage}
    Finally, the bottom panel of Figure \ref{fig:photon_counts_correlations} shows the total signal including all sources using a perturbed instrument. Several important observations can be made here. The total signal is almost an order of magnitude higher than in the shot-noise case. These additional counts are mostly caused by stellar photons and result from the degradation of the null or the change of the instrument response for on-axis sources, respectively. We have already seen this in Figure \ref{fig:intensity_responses}, where the central white fringe of the ideal response (left) is no longer preserved in the perturbed case (right). Figure \ref{fig:intensity_response_star} illustrates the same effect on the scale of the stellar extent (indicated by the dark gray circle). 
    The top row shows the response of an ideal instrument (left) and a perturbed one (right) at a given instance in time. While the ideal response is perfectly symmetric and white in the center, the perturbed response shows asymmetries and a non-zero response in the center, allowing more stellar photons to leak onto the detector. This is even more apparent in the two bottom panels of Figure \ref{fig:intensity_response_star}, which show the RMS of the instrument response over a full rotation of the array. The extent of the null in the perturbed instrument is significantly reduced within the stellar disk.
    
\begin{figure}
        \centering

            \includegraphics[width=0.75\linewidth]{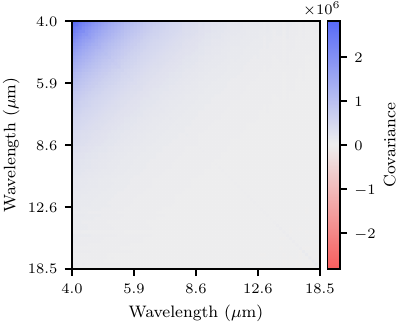}
        \caption{Covariance for the bottom panel of Figure \ref{fig:photon_counts_correlations}, confirming strong correlations between wavelength bins predominantly at shorter wavelengths with a falloff towards longer wavelengths. Note that in this case correlations are mostly positive. However, this is not generally required and depends on the instrument properties.}
        \label{fig:covariance_matrix}
    \end{figure}

    \paragraph{Spectral Correlations}
    A striking feature in the bottom panel of Figure \ref{fig:photon_counts_correlations} are the vertical stripes that fall off in their intensity from top to bottom. This indicates the presence of errors, which are strongly correlated across different wavelength bins. This is highlighted in Figure \ref{fig:covariance_matrix}, which shows the covariance of the wavelength channels in the bottom panel of Figure \ref{fig:photon_counts_correlations}.
    The falloff of the counts towards longer wavelengths is a consequence of the shape of the stellar SED, which is lower at longer wavelengths, but also of the $\lambda^{-1}$ dependence of the phase perturbation terms in Equation (\ref{eq:phase_pert}). The key aspect of this study is to understand the impact of these correlations on signal extraction and techniques to mitigate them.

    \paragraph{Temporal Correlations}
    Less striking, but also visible in the bottom panel of Figure \ref{fig:photon_counts_correlations} are temporal correlations. Contrary to the ideal signal (center panel), where the counts at different time steps are independent of each other (shot noise), the vertical stripes in the bottom panel show a dependence on the values of the neighboring stripes. As mentioned earlier, an explicit treatment of temporal correlations is outside of the scope of this study.

    \subsection{Synthetic Data Generation With PHRINGE}
    To enable the generation of synthetic data that includes instrumental errors, we have developed \textsc{PHRINGE}\footnote{\url{https://github.com/pahuber/PHRINGE}}, a PHotoelectron counts generatoR for nullING intErferometers. It offers state-of-the-art, GPU-accelerated generation of synthetic data of exoplanetary systems recorded by space-based nullers and has been designed to be flexible in the instrument architectures, noise properties and astrophysical scenes it can model. It is also capable of exporting simulated data in the new nulling interferometry data standard, NIFITS, which is currently being developed and implemented in the Python package nifits\footnote{\url{https://github.com/rlaugier/nifits}}. This enables \textsc{PHRINGE} to effortlessly interface with future nulling data processing pipelines. It is available as an open-source Python package and is also used within \textsc{LIFEsimMC} (see Section \ref{subsec:lifesimmc}), our new Monte Carlo-based end-to-end simulator for LIFE. In fact, \textsc{PHRINGE} leverages Monte Carlo simulations to explicitly calculate the photon counts, $N_j$, as described in Equation (\ref{eq:photon_counts}), at each time and wavelength step. To this end, it generates random values for the instrument perturbations according to their underlying distributions and propagates them through the equations. 

%
%

    \section{Data processing and signal extraction}\label{sec:data_processing_signal_extraction}
    
    \subsection{Data Whitening}\label{subsec:whitening}
\begin{figure}
        \centering
        \includegraphics[width=1\linewidth]{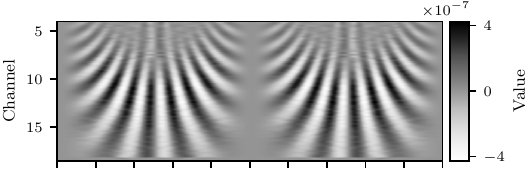}\vspace{0.1cm}
        \includegraphics[width=1\linewidth]{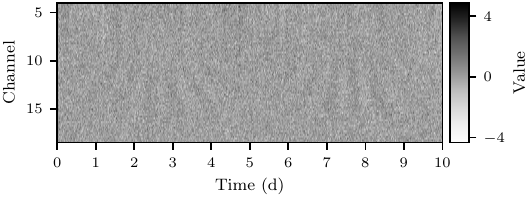}
        \caption{Illustration of the effect of the whitening transformation on the planetary model (top) and full data (bottom). The correlations visible in the bottom panel of Figure \ref{fig:photon_counts_correlations} have been removed and the noise appears completely white, whereas the planetary signal is only slightly changed, as expected. Note that the panels now show unitless values and do no longer represent physical counts.}
        \label{fig:whitening}
    \end{figure}

    The data recorded in the differential output of the double Bracewell nuller have been illustrated in the three panels of Figure \ref{fig:photon_counts_correlations} as matrices of shape $n_\lambda \times n_t$. Here, $n_\lambda$ is the number of wavelength bins and $n_t$ the number of time steps. In general, this data is a three dimensional $n_\text{diff} \times n_\lambda \times n_t$ data cube, 
    \begin{equation}\label{eq:data_cube}
        \begin{aligned}
            \boldmat{D}&=(D_{m\tau l})\\
            &\equiv\Delta N_m(t_\tau, \lambda_l),
        \end{aligned}
    \end{equation}
    which we obtain by evaluating Equation (\ref{eq:differential_null}) for all $m \in (1, n_\text{diff})$ differential outputs, discrete time steps $t_\tau$ with $\tau\in (1, n_t)$ and discrete wavelengths $\lambda_l$ with $l\in(1, n_\lambda)$. In this section, we treat this data cube as a one-dimensional vector, $\boldvec{\mathrm{y}}'$, containing $n_\text{diff}n_\lambda$ random variables, of which we acquire $n_t$ samples during an observation. We can express $\boldvec{\mathrm{y}}'$ also as the sum of the planet signal or model, $\boldvec{\mathrm{x}}'$, and the noise, $\boldvec{\varepsilon}'$, 
    \begin{equation}\label{eq:model_fit}
        \boldvec{\mathrm{y}}'=\boldvec{\mathrm{x}}'+\boldvec{\varepsilon}'.
    \end{equation}
    Here, we make the assumption that, when the model, $\boldmat{x}'$, matches the data, $\boldmat{y}'$, then $\boldmat{\varepsilon}'\sim \mathcal{N}(0, \boldmat{\Sigma})$, where $\boldmat{\Sigma}$ is the covariance of the noise \citep{laugier_asgardnott_2023}, in the following referred to as the \textit{instrumental error covariance}. More precisely, $\boldmat{\Sigma}$ is an $n_\text{diff}n_\lambda \times n_\text{diff}n_\lambda$ matrix containing the covariance of the elements of $\boldvec{\mathrm{y}}'$. As we have seen in Figures \ref{fig:photon_counts_correlations} and \ref{fig:covariance_matrix}, $\boldvec{\mathrm{y}}'$ generally contains strong noise contributions, which are highly correlated and not independent. This not only complicates signal extraction (see Section \ref{subsec:parameter_estimation}), but also violates the assumptions used in the hypothesis tests (see Section \ref{subsec:hypothesis_testing}). We resolve this by performing data whitening as suggested by \cite{ceau_kernel_phase_2019} by multiplying the whitening matrix $\boldmat{W}$ by the unwhitened vectors,
    \begin{gather}
        \boldvec{\mathrm{y}}\equiv\boldmat{W}\boldvec{\mathrm{y}}',\label{eq:whitening_w1}\\
        \boldvec{\mathrm{x}}\equiv\boldmat{W}\boldvec{\mathrm{x}}'.\label{eq:whitening_w2}
    \end{gather}
    Note that there are different ways to define the whitening matrix $\boldmat{W}$. Common choices include zero-phase component analysis (ZCA) whitening with $\boldmat{W}=\boldmat{\Sigma}^{-\frac{1}{2}}$ and principal component analysis (PCA) whitening with $\boldmat{W}=\boldmat{\Lambda}^{-\frac{1}{2}}\boldmat{U}^T$, where $\boldmat{\Sigma}=\boldmat{U}\boldmat{\Lambda}\boldmat{U}^T$ is the eigendecomposition of $\boldmat{\Sigma}$ \citep{kessy_optimal_2018}. PCA is predominantly useful when a reduction in dimensionality is required. If no dimensionality reduction is performed, PCA and ZCA yield equivalent results in terms of signal extraction and hypothesis testing. Therefore, and to comply with \cite{ceau_kernel_phase_2019}, we choose to continue with ZCA whitening in this work. We thus obtain the whitened data and model by inserting $\boldmat{W}=\boldmat{\Sigma}^{-\frac{1}{2}}$ into Equations (\ref{eq:whitening_w1}) and (\ref{eq:whitening_w2}):
    \begin{gather}
        \boldvec{\mathrm{y}}\equiv\boldmat{\Sigma} ^{-\frac{1}{2}}\boldvec{\mathrm{y}}',\label{eq:whitening_1}\\
        \boldvec{\mathrm{x}}\equiv\boldmat{\Sigma} ^{-\frac{1}{2}}\boldvec{\mathrm{x}}'.\label{eq:whitening_2}
    \end{gather}
    This leads to a new model of our data,
    \begin{equation}\label{eq:model_fit_white1}
        \boldvec{\mathrm{y}}=\boldvec{\mathrm{x}}+\boldvec{\varepsilon},
    \end{equation}
    where now $\boldvec{\varepsilon}\sim \mathcal{N}(0, \boldmat{I})$. Figure \ref{fig:whitening} illustrates the data and the model after the whitening transformation. 
    The impact on the planetary signature is minimal, while the total signal has changed significantly. Notably, the spectral correlations have been removed and the noise appears to be white. Indeed, the planetary signature has become slightly visible in the whitened data.\par
    
    In practice, obtaining $\boldvec{\Sigma}=\boldmat{Cov}\left(\boldvec{\varepsilon}'\right)$ can be challenging, as a measurement of a planet only provides $\boldvec{\mathrm{y}}'=\boldvec{\mathrm{x}}'+\boldvec{\varepsilon}'$, that is, the sum of planet signal and noise, and not noise alone. For simplicity, we work here under the hypothesis that acquisitions on a calibration star, $\boldvec{\mathrm{y}}_\text{ref}'=\boldvec{\varepsilon}_\text{ref}'$, provide us with a relevant estimate of the covariance of errors on the science target,
    \begin{equation}\label{eq:noise_covariance}
        \boldvec{\Sigma}_\text{ref}=\boldmat{Cov}\left(\boldvec{\mathrm{y}}_\text{ref}'\right).
    \end{equation}
    We further discuss this approach and alternatives in Section \ref{subsec:dependence_on_instrument_error_model}.
    

    \subsection{Statistical Hypothesis Testing}\label{subsec:hypothesis_testing}
    Statistical hypothesis testing is used to make informed decisions about whether a signal of interest is present in our data. In particular, given the whitened data and model from above, the goal is to differentiate between the two statistical hypotheses,
    \begin{equation}\label{eq:hypotheses}
        \begin{cases}
            \mathcal{H}_0 : \boldmat{y}=\boldmat{\varepsilon},\\
            \mathcal{H}_1 : \boldmat{y}=\boldmat{x} + \boldmat{\varepsilon},
        \end{cases}
    \end{equation}
    where $\mathcal{H}_0$ is the null hypothesis, under which no planetary signal is present, and $\mathcal{H}_1$ is the alternative hypothesis, where a signal is present. We follow the approach of \cite{laugier_asgardnott_2023} and use the statistical tests that \cite{ceau_kernel_phase_2019} describe for detections in kernel phases for detection tests on our whitened data, $\boldmat{y}$. We briefly describe the Neyman-Pearson and energy detector tests in Sections \ref{subsubsec:neyman} and \ref{subsubsec:energy}, and refer the reader to \cite{laugier_asgardnott_2023} for a more detailed description thereof.

    \subsubsection{Neyman-Pearson Test}\label{subsubsec:neyman}
    The likelihood ratio or Neyman-Pearson test \citep{neyman_pearson_1933} is a suitable test for the problem defined in Equation (\ref{eq:hypotheses}) and requires knowledge of the model, $\boldmat{x}$ \citep{ceau_kernel_phase_2019}. It can be regarded as an upper bound to our detection problem. It is based on the likelihood function for Gaussian white noise, which reads \citep{scharf_matched_1994}
    \begin{equation}\label{eq:likelihood_flattened}
        \mathcal{L}(\boldmat{x}; \boldmat{y})=(2\pi)^{-n_\text{diff}n_t n_\lambda /2}\exp\left(-\frac{1}{2}(\boldmat{x}-\boldmat{y})^T(\boldmat{x}-\boldmat{y})\right).
    \end{equation}
    Calculating the ratio of the likelihoods under the alternative and the null hypothesis, $\mathcal{L(\boldmat{x};\boldmat{y}})/\mathcal{L}(\boldmat{0};\boldmat{
    y})$, we can express the test statistic of the Neyman-Pearson test as
    \begin{equation}\label{eq:t_np}
        T_\text{NP}(\boldmat{y}, \boldmat{x})=\boldmat{y}^T\boldmat{x}\;\substack{\mathcal{H}_1\\\displaystyle\gtrless\\\mathcal{H}_0}\;\xi_\text{NP}.
    \end{equation}
    This test statistic is distributed as \citep{ceau_kernel_phase_2019}
    \begin{equation}
        \begin{cases}
            \mathcal{H}_0 : T_\text{NP}(\boldmat{y}, \boldmat{x})\sim \mathcal{N}\left(0, \boldmat{x}^T\boldmat{x}\right),\\
            \mathcal{H}_1 : T_\text{NP}(\boldmat{y}, \boldmat{x})\sim \mathcal{N}\left(\boldmat{x}^T\boldmat{x}, \boldmat{x}^T\boldmat{x}\right).
        \end{cases}
    \end{equation}
    The threshold, $\xi_\text{NP}$, upon which either the null or alternative hypothesis is accepted, can be determined by
    \begin{equation}\label{eq:xi_np}
        \xi_\text{NP} = \mathcal{F}_{\mathcal{N}(0, \boldmat{x}^T\boldmat{x})}^{-1}\left(1-P_\text{FA}\right),
    \end{equation}
    where $\mathcal{F}_{\mathcal{N}(0, \boldmat{x}^T\boldmat{x})}^{-1}$ is the inverse cumulative distribution function (CDF) of the normal distribution with $\mu=0$ and $\sigma^2=\boldmat{x}^T\boldmat{x}$ and $P_\text{FA}$ is the probability of false alarm. 
    In general, the true model, $\boldmat{x}$, is not known a priori and we rely on an estimator of the model, $\hat{\boldmat{x}}$. We describe the use of numerical maximum likelihood estimation to obtain this estimator in Section \ref{subsec:parameter_estimation}.
    
    \subsubsection{Energy Detector Test}\label{subsubsec:energy}
    An alternative test is given by the energy detector test. It assumes no prior knowledge whatsoever, setting $\hat{\boldmat{x}}=\boldmat{y}$ \citep{ceau_kernel_phase_2019}. This represents the \enquote{worst-case} scenario and can thus be viewed as a lower bound for the detection performance. Its test statistic is defined as
    \begin{equation}
        T_\text{ED}(\boldmat{y})\equiv\boldmat{y}^T\boldmat{y} \;\substack{\mathcal{H}_1\\\displaystyle\gtrless\\\mathcal{H}_0}\;\xi_\text{ED},
    \end{equation}
    and distributed as \citep{ceau_kernel_phase_2019}
    \begin{equation}
        \begin{cases}
            \mathcal{H}_0: T_\text{ED}(\boldmat{y}) \sim \chi^2_p(\delta^2=0),\\
            \mathcal{H}_1: T_\text{ED}(\boldmat{y}) \sim \chi^2_p(\delta^2 = \boldmat{x}^T\boldmat{x}),
        \end{cases}
    \end{equation}
    where $\chi^2_p$ is a chi-square distribution with $p=n_\text{diff}n_\lambda$ degrees of freedom and non-centrality parameter $\delta$.
    The threshold, $\xi_\text{ED}$, is given by
    \begin{equation}
        \xi_\text{ED} = \mathcal{F}_{\chi^2_p(0)}^{-1}\left(1-P_\text{FA}\right),
    \end{equation}
    where $\mathcal{F}_{\chi^2_p(0)}^{-1}$ is the inverse CDF of $\chi^2_p(\delta^2 =0)$.
    

    \subsection{Signal Extraction}\label{subsec:parameter_estimation}
    The goal of signal extraction is to obtain an estimator of the planetary position, $\hat{\boldvec{\alpha}}_\text{p}$ and SED, $\hat{F}_\text{p}$, which also allows for an estimation of its temperature and radius, e.g., by fitting a black body spectrum to $\hat{F}_\text{p}$. This is equivalent to finding an estimator, $\hat{\boldmat{x}}$, which can then also be used for the Neyman-Pearson test.
    We can rewrite out Equation (\ref{eq:likelihood_flattened}) explicitly in terms of the planetary properties $F_\text{p}$ and $\boldvec{\alpha}_\text{p}$ by using Equations (\ref{eq:template}) and (\ref{eq:data_cube}), to obtain
    \begin{equation}\label{eq:likelihood}
        \begin{aligned}
            &\mathcal{L}\left(\boldvec{F}_\text{p}, \boldvec{\alpha}_\text{p};\boldmat{D}\right)\propto\\
            &\exp\left(-\frac{1}{2}\sum_{m=1}^{n_\text{diff}}\sum_{\tau=1}^{n_t}\sum_{l=1}^{n_\lambda}\left(F_{\text{p}}(\lambda_l)T_{\text{p}, m}\left(t_\tau, \lambda_l, \boldvec{\alpha}_\text{p}\right)-D_{m \tau \ell}\right)^2\right).
        \end{aligned}
    \end{equation}
    Note that for convenience, we collect the fluxes $F_\text{p}$ in a vector $\boldvec{F}_\text{p}$, of length $n_\lambda$. For a given instrument, the planetary signal depends only on the planet's SED, $\boldvec{F}_\text{p}$, and its position,
    $\boldvec{\alpha}_\text{p}$.\par
    
    A common approach to finding $\hat{\boldvec{F}}_\text{p}$ and $\hat{\boldvec{\alpha}}_\text{p}$ is given by the maximum likelihood estimation \citep{thiebaut_maximum_2005, mugnier_data_2006, dannert_large_2022}. This entails maximizing the likelihood in Equation (\ref{eq:likelihood}) with respect to $\boldvec{F}_\text{p}$ and $\boldvec{\alpha}_\text{p}$ or, equivalently minimizing the negative log-likelihood function, 
    \begin{equation}\label{eq:neg_log_likelihood}
        \ell\left(\boldvec{F}_\text{p}, \boldvec{\alpha}_\text{p};\boldmat{D}\right)=\frac{1}{2}\sum_{m=1}^{n_\text{diff}}\sum_{\tau=1}^{n_t}\sum_{l=1}^{n_\lambda}\left(F_{\text{p}}(\lambda_l)T_{\text{p}, m}\left(t_\tau, \lambda_l, \boldvec{\alpha}_\text{p}\right)-D_{m \tau \ell}\right)^2
    \end{equation}
    with respect to these variables. Written out explicitly, the estimators for the SED and position are then given by
    \begin{equation}\label{eq:arg_min_log}
        \left(\hat{\boldvec{F}_\text{p}}, \hat{\boldvec{\alpha}}_\text{p}\right)=\arg \min_{(\boldvec{F}, \boldvec{\alpha})} \ell\left(\boldvec{F}_\text{p}, \boldvec{\alpha}_\text{p};\boldmat{D}\right).
    \end{equation}
    Typically, Equation (\ref{eq:arg_min_log}) is solved using a matched filter approach \citep{thiebaut_maximum_2005, mugnier_data_2006, dannert_large_2022}. First, the template function that best describes the data is found by evaluating Equation (\ref{eq:neg_log_likelihood}) on a grid of possible planet positions. The global minimum of the map created in this way indicates the most likely planet position, yielding the corresponding template function. Given this template function, Equation (\ref{eq:arg_min_log}) can be solved analytically \citep{dannert_large_2022}.\par
    
    In this study, we refrain from this analytical approach and solve Equation (\ref{eq:arg_min_log}) numerically by performing a least-squares minimization based on the Levenberg-Marquardt algorithm, as implemented in the Python package lmfit \citep{newville_2015_11813}. This has the inherent benefit of providing us directly with an estimate of the full covariance of the estimated parameters $\hat{\boldvec{F}}_\text{p}$ and $\hat{\boldvec{\alpha}}_\text{p}$, an approach that could also be applied to real single-epoch data. Contrary to this, the analytical method relies on repeated simulation of the same observation and repeated extraction of the parameters to calculate the covariance from the statistics of the results \citep{dannert_large_2022}.
    To explain the covariance estimation of the numerical approach, we first concatenate our parameter vectors, $\boldvec{F}_\text{p}$, $\boldvec{\alpha}_\text{p}$, into a single vector of length $n_\lambda + 2$,
    \begin{equation}
        \boldmat{\Theta}\equiv \left[F_{\text{p}, 1},\hdots, F_{\text{p}, n_\lambda}, \alpha_{\text{p}}, \beta_{\text{p}}\right]^T.
    \end{equation}
    We write the negative log-likelihood function in Equation (\ref{eq:neg_log_likelihood}) also as a function of $\boldvec{\Theta}$, as $\ell\left(\boldvec{\Theta};\boldmat{D}\right)$. Using the finite difference approximation \citep{newville_2015_11813} we calculate the Hessian, $\boldmat{H}_\ell$, of the negative log-likelihood function, which is a $(n_\lambda +2)\times (n_\lambda +2)$ square matrix with elements
        \begin{equation}
        H_{uv}\left(\boldvec{\Theta}\right)=\frac{\partial^2 \ell\left(\boldvec{\Theta};\boldmat{D}\right)}{\partial \Theta_{ u} \,\partial \Theta_{v}}.
    \end{equation}
    Here, $\Theta_{u}$ and $\Theta_{v}$ are the $u$th and $v$th elements of $\boldvec{\Theta}$, respectively. An estimate of the covariance of $\boldvec{\Theta}$, is then given by the $(n_\lambda + 2)\times (n_\lambda + 2)$ matrix,
    \begin{equation}\label{eq:covariance_matrix}
        \hat{\boldvec{\Sigma}}_{\boldvec{\Theta}}\approx 2 \boldmat{H}_\ell^{-1}\left(\hat{\boldvec{\Theta}}\right),
    \end{equation}
    where $\boldmat{H}_\ell^{-1}$ is the inverse of the Hessian of the negative log-likelihood function evaluated at $\hat{\boldmat{\Theta}}$ \citep{newville_2015_11813}. In the following, we refer to $\hat{\boldvec{\Sigma}}_{\boldvec{\Theta}}$ as the \textit{spectral covariance}. An estimate of the standard deviations of the SED, $\hat{\boldvec{F}}_\text{p}$, and position, $\hat{\boldvec{\alpha}}_\text{p}$, can then be extracted from the diagonal elements of Equation (\ref{eq:covariance_matrix}). Note that the spectral covariance, $\hat{\boldvec{\Sigma}}_{\boldvec{\Theta}}$, is calculated in a different context and must not be confused with the instrumental error covariance, $\boldvec{\Sigma}$, from Equation (\ref{eq:noise_covariance}), which we have used for the data whitening.

    \subsection{End-to-End Simulations With LIFEsimMC}\label{subsec:lifesimmc}
    We have also developed \textsc{LIFEsimMC}\footnote{\url{https://github.com/pahuber/LIFEsimMC}}, a new Monte Carlo-based end-to-end simulator for LIFE that is publicly available as a Python package. It uses \textsc{PHRINGE} internally for the data generation and thus offers the same flexibility in terms of instrument architectures, noise properties, and astrophysical scenes. It is based on a modular pipeline architecture and extends the data generation by a suite of processing and signal extraction tools, connecting the astrophysical and instrumental input to the extracted planetary properties. These include in particular data whitening, detections tests, correlation maps, and maximum-likelihood estimation of the planetary SED and position, as partially detailed in this section. \textsc{LIFEsimMC} has been designed to support the requirement derivation process for the LIFE mission and is a natural extension of the previously introduced tool \textsc{LIFEsim} \citep{dannert_large_2022}, which is not able to account for instrumental noise. Currently, \textsc{LIFEsimMC} can only be used to simulate single-epoch observations of individual systems, but we intend to extend its capabilities to population-based observations in future work.

    

%
%

    \section{Application \& results}\label{sec:results}
    We use \textsc{PHRINGE} and \textsc{LIFEsimMC} to apply the framework outlined in Sections \ref{sec:data} and \ref{sec:data_processing_signal_extraction} to the reference case of an observation of an Earth twin around a Sun twin at \SI{10}{pc}. All mission and astrophysical parameters that are used are specified in Table \ref{tab:mission_parameters}. We consider different scenarios corresponding to different levels of instrumental perturbations, ranging from no perturbations, i.e., an ideal instrument, to optimistic and pessimistic levels (see Table \ref{tab:scenarios}), and examine the impact of whitening on the respective performances.
    \begin{table}
    \begin{threeparttable}[b]
        \caption{Instrument perturbation values corresponding to the ideal, optimistic, and pessimistic scenarios used in this work.}
        \label{tab:scenarios}
        \setlength{\tabcolsep}{3pt} 
        \begin{small} 
        \begin{tabularx}{\linewidth}{l>{\centering\arraybackslash}p{2cm}>{\centering\arraybackslash}p{2cm}>{\centering\arraybackslash}p{2cm}}\toprule
             Scenario & \multicolumn{1}{c}{$\mathrm{RMS}(\delta A_k)$ (\%)} & \multicolumn{1}{c}{$\mathrm{RMS}(\delta p_k)$ (nm)} & \multicolumn{1}{c}{$\mathrm{RMS}(\delta \theta_k)$ (rad)} \\\midrule
             Ideal & 0.0 & 0.0 & 0.000\\
            Optimistic & 0.1\tnote{a} & 1.5\tnote{a} & 0.001\tnote{a}\\
            Pessimistic & 1.0 & 15.0 & 0.010\\
            \bottomrule
        \end{tabularx}
        \end{small}
        \begin{tablenotes}
            \item [a] Baseline values as proposed by \cite{lay_systematic_2004}.
        \end{tablenotes}
    \end{threeparttable}
\end{table}
    The optimistic perturbation levels are based on the values from \cite{lay_systematic_2004}, which we have scaled by a factor of ten to represent the pessimistic levels. To make results for unwhitened and whitened data numerically comparable in the following analysis, we apply the whitening transformation described in Equations (\ref{eq:whitening_1}) and (\ref{eq:whitening_2}) also for the ideal and unwhitened cases, using, however, only the diagonal of the instrumental error covariance matrix, $\boldvec{\Sigma}$, standardizing the data with the variance.\par
    \begin{figure*}
        \centering
        \includegraphics[width=\linewidth]{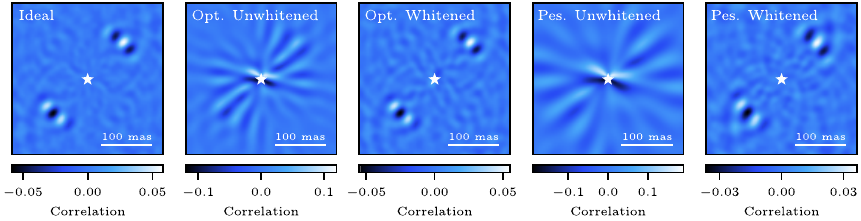}\vspace{0.2cm}
        \caption{Correlation maps showing the correlation of the data with the planetary templates, $\boldvec{\mathrm{y}}^T\boldvec{\mathrm{x}}_I/\sqrt{\boldvec{\mathrm{x}}_I^T\boldvec{\mathrm{x}}_I}$, where $\boldvec{\mathrm{x}}_I$ describes a model with unit SED. The ideal scenario shows the clearest sign of the planet in the upper right corner. The maps based on unwhitened data show strong artifacts and do no longer allow for a detection of the planet in the map. The capability to locate the planet at its correct position is improved significantly for the whitened scenarios.}
        \label{fig:pseudo_image_comparison}
    \end{figure*}
    As a first figure of merit, so-called \textit{correlation maps} are shown in Figure \ref{fig:pseudo_image_comparison} for the three scenarios with and without whitening. They are calculated as $\boldvec{\mathrm{y}}^T\boldvec{\mathrm{x}}_I/\sqrt{\boldvec{\mathrm{x}}_I^T\boldvec{\mathrm{x}}_I}$, where $\boldvec{\mathrm{x}}_I$ corresponds to the model at a given pixel and has unit SED. Alternatively, one could also consider correlating the data with the model of a blackbody spectrum. The correlation map of the ideal scenario can be considered as the benchmark here and shows a clear maximum at the correct position of the planet in the upper right corner. The adjacent minima, as well as the other minimum and maxima on the opposite side of the star, are part of the expected planetary signature and arise because of symmetries. Both maps corresponding to unwhitened data show strong artifacts and no longer allow for a clear identification of the planet. Indeed, multiple patches exist that could be mistaken for the signature of a planet. 
    In both cases, whitening practically restores the appearance and hence the usability of these maps to the one of the ideal map, showing again clearly the planetary signature, albeit with a slight increase in background noise and a reduction of the correlation values for pessimistic levels. In Appendix \ref{app:subsec_m_stars}, we repeat this analysis for an M star system, as LIFE is also capable of detecting and characterizing planets around such stars \citep{quanz_large_2022}.\par
    \begin{figure*}
        \includegraphics[width=\linewidth]{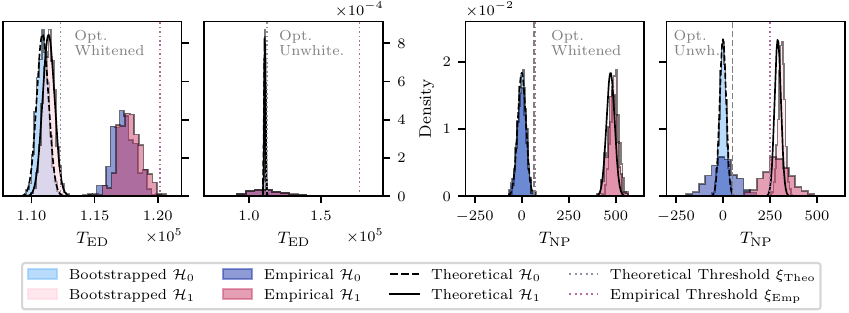}
        \caption{Distribution of the test statistics, $T_\text{ED}$ (left two plots) and $T_\text{NP}$ (right two plots), under the null hypothesis, $\mathcal{H}_0$, and alternative hypothesis, $\mathcal{H}_1$, for the optimistic scenario with and without whitening. A $P_\text{FA}=0.00135$ corresponding to a $3\sigma$ detection is assumed. The empirical results are calculated for a sample of 500 simulated observations of our target system. The bootstrapped results are sampled from a multivariate normal distribution following the considered instrumental error covariance, $\boldmat{\Sigma}_\text{ref}$.} 
        \label{fig:tests}
    \end{figure*}
    The effect of whitening is further investigated by comparing the distributions of the test statistics, $T_\text{ED}$ and $T_\text{NP}$, drawn from 500 realizations of the observation. As shown in Figure \ref{fig:tests}, whitening reduces the spread of their distributions and brings them closer to their theoretical probability density functions, improving the sensitivity and usability of the tests. However, for the case of $T_\text{ED}$, we still observe that the empirical distribution differs from the theory. This can be attributed to two effects neglected by our approach: the slight deviation from Gaussian distribution in the differential observable as identified by \cite{dannert_consequences_2025} and the temporal correlations of instrumental errors. This is verified through bootstrapping by drawing random multivariate normal samples following the considered covariance estimation, for which the test statistic matches expectations. Accounting for these effects will be the topic of future work. Note that for the chosen astrophysical, instrumental, and noise properties the energy detector test is not useful for claiming detections, as it can hardly distinguish the distributions under $\mathcal{H}_0$ and $\mathcal{H}_1$.\par
\begin{figure}
        \centering
        \includegraphics[width=\linewidth]{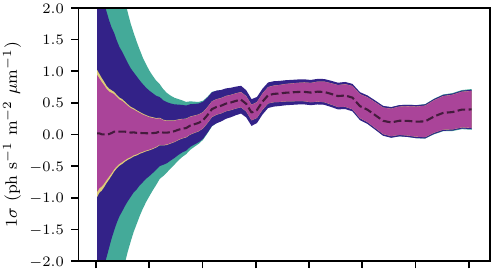}\vspace{-0.2cm}
        \begin{flushright}
        \includegraphics[width=0.944\linewidth]{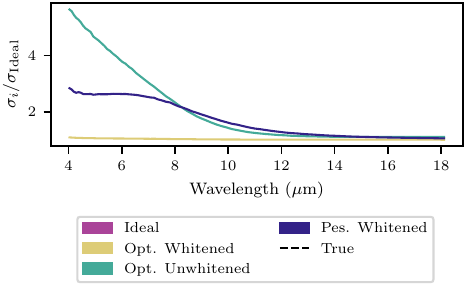}
        \end{flushright}
        \caption{Comparison of the $1\sigma$ uncertainties of the estimated planetary SED derived from the diagonal terms of Equation (\ref{eq:covariance_matrix}) for the different scenarios. To ensure comparability, it is assumed here that the correct planet position is known a priori, even for the optimistic unwhitened case. \textit{Top:} The black dashed line shows the true planetary spectrum and the shaded areas indicate the uncertainties obtained from averaging over fifty simulated observations. Significant differences are seen below $\sim \SI{10}{\mu m}$. \textit{Bottom:} The $1\sigma$ uncertainties of the optimistic whitened, optimistic unwhitened, and pessimistic whitened cases, normalized by the ideal uncertainty. Results for the pessimistic unwhitened scenario are not shown, as the algorithm was not able to converge.}
        \label{fig:ref_sed_comparison}
    \end{figure}

    Figure \ref{fig:ref_sed_comparison} (top) shows the $1\sigma$ uncertainty or standard deviation of the estimated planetary SED after signal extraction. 
    Here, it is assumed that the true planetary position is known before the SED is estimated, allowing us to directly assess the effect of whitening on the uncertainty. The standard errors of the optimistic whitened case are similar in magnitude to the ideal ones, whereas for the pessimistic whitened and optimistic unwhitened cases they show a significant increase below $\sim \SI{10}{\mu m}$. At $\SI{4}{\mu m}$, for instance, we obtain an increase of $\sim 3$ or $\sim 6$, respectively, compared to the ideal case (see Figure \ref{fig:ref_sed_comparison} (bottom)), potentially impacting the results of atmospheric retrievals at short wavelengths.
    \begin{figure}
        \centering
        \includegraphics[width=\linewidth]{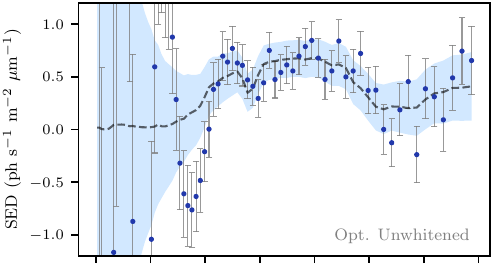}\vspace{0.1cm}
        \includegraphics[width=\linewidth]{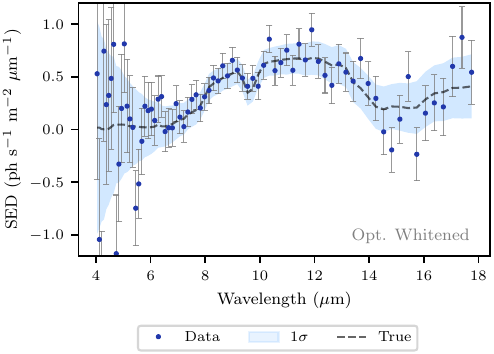}
        \caption{Estimated SED for the optimistic scenario without data whitening (top) and with data whitening (bottom), assuming the true planet position is known a priori. Apart from the larger uncertainties in the non-whitened case, the SED values at shorter wavelengths show prominent systematic errors. These systematics are significantly reduced after the whitening.}
        \label{fig:sed_systematic_error}
    \end{figure}
    Figure \ref{fig:sed_systematic_error} shows two examples of the estimated SED for the optimistic scenario with and without whitening. The unwhitened estimates show strong systematic errors at shorter wavelengths, an effect that has already been discussed in \citep{matsuo_large_2023}. These are reduced after whitening. We will discuss these systematic errors further in Section \ref{subsec:corr_SED}.\par

    \section{Discussion}\label{sec:discussion}

    \subsection{Implications for Detection and Characterization}
    The results presented in Section \ref{sec:results} suggest that a moderately perturbed interferometer (optimistic scenario) performs similarly to an ideal instrument in both detection and characterization. While this is promising, the significance of this result is tempered by the assumptions about the relatively simple wavelength dependence of the instrumental perturbations (see Equation (\ref{eq:phase_pert})) and the availability of the instrumental error covariance through a calibration observation (see Section \ref{subsec:dependence_on_instrument_error_model}). The results have also shown that instrumental errors in the form of instability noise predominantly affect the characterization of the planetary atmosphere below $\sim \SI{10}{\mu m}$, given a more strongly perturbed instrument (pessimistic scenario). This may affect the retrieval of certain atmospheric constituents, whose signatures are predominately expected at those wavelengths, such as methane \citep{konrad_large_2022}.
    
    \subsection{Obtaining the Instrumental Error Covariance}\label{subsec:dependence_on_instrument_error_model}
    As outlined in Section \ref{subsec:whitening}, the prerequisite for data whitening is knowledge of the instrumental error covariance, $\boldvec{\Sigma}=\boldmat{Cov}\left(\boldvec{\varepsilon}'\right)$. In this work, we have limited ourselves to the best-case scenario of having an ideal reference star (in the sense of having properties identical to those of the target star), as we have focused on the exploration of the theoretical implications of the whitening rather than the direct applicability to on-sky observations. In practice, different approaches will be required to obtain a good model of the instrumental errors. One approach that makes use of simulations of non-ideal reference star observations is followed by \cite{laugier_asgardnott_2023}. They model the part of the noise covariance corresponding to the instrumental noise as
    \begin{equation}
        \boldvec{\Sigma}_\text{instrumental} \approx \frac{F_\star}{F_\text{ref}} \boldvec{\Sigma}_\text{ref, MC},
    \end{equation}
    where $F_\star$ is the total flux of the star of the current observation, $F_\text{ref}$ is the total flux of the reference star, and $\boldvec{\Sigma}_\text{ref, MC}$ is the noise covariance, which they evaluate for different pointing directions using MC simulations. This approach allows them to build a parametric model of the instrument noise, which can then be used for data whitening. Another step in the direction of using calibrator stars is made by \cite{ireland_phase_2013}. The author compares several calibration strategies, highlighting their biases, and describes the POISE (phase observationally independent of systematic errors) technique, which may also be relevant in the context of obtaining a suitable instrumental error covariance. In addition to calibrator stars, internal calibrator lasers may also be a viable option that should be explored in future work.\par
    As mentioned earlier in Section \ref{sec:intro}, alternative approaches exist that do not rely on calibration observations. For example, \cite{lay_removing_2006} and \cite{matsuo_large_2023} have used data-driven approaches that make assumptions about the shape of the spectral correlations (see bottom panel of Figure \ref{fig:photon_counts_correlations}) to fit or extrapolate them across the wavelength bins before finally subtracting them from the data. Although both approaches are successful in improving instrument performance, neither ensures that the test statistic follows a controlled distribution, as assumed in signal extraction and hypothesis testing, a property that is, in principle, inherent to data whitening.\par
    Recent progress has been made by \cite{flasseur_shrinkage_2024} in improving the estimates made from the data with analytical or numerical models. Such numerical models could be derived from simulators such as \textsc{PHRINGE}. The incorporation of the ZCA whitening transformation done here into a data-driven approach could be another step toward a suitable treatment of instrumental errors that is applicable to on-sky data. In this regard, it is also interesting to look at methods applied in classical high-contrast imaging of exoplanets, where the estimation and successive subtraction of the stellar point spread function from the data is well established \citep[e.g.][]{Meshkat_2014, gomez_low_2016}. A precise and accurate estimation of the covariance of an observation is ubiquitous in astronomical observations.  More work is needed to develop more practical approaches, which we leave for future research.

    \subsection{Using Spectral Covariance in Atmospheric Retrievals}\label{subsec:corr_SED}
\begin{figure}
        \centering
        \includegraphics[width=0.8\linewidth]{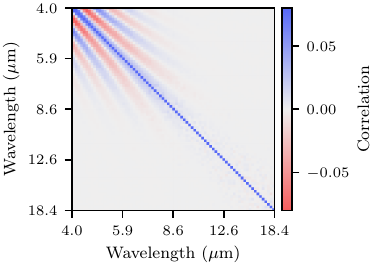}
        \caption{Illustrative example of the correlation matrix corresponding to the covariance matrix defined in Equation(\ref{eq:covariance_matrix}), averaged over 100 runs. The color map is capped at $\pm 0.08$ to increase the visibility of the smaller features. The values on the diagonal are one. This pattern is similar to the covariance of the planetary signal as shown in Figure \ref{fig:photon_counts_correlations} (top).}
        \label{fig:correlation_comparison}
    \end{figure}

    As anticipated in Section \ref{subsec:parameter_estimation}, Equation (\ref{eq:covariance_matrix}) provides a numerical estimate of the spectral covariance, i.e. the covariance of the estimated planetary SED. Figure \ref{fig:correlation_comparison} shows the corresponding correlation matrix for the optimistic scenario after whitening, averaged over 100 runs. 
    \begin{figure}
        \centering
        \includegraphics[width=\linewidth]{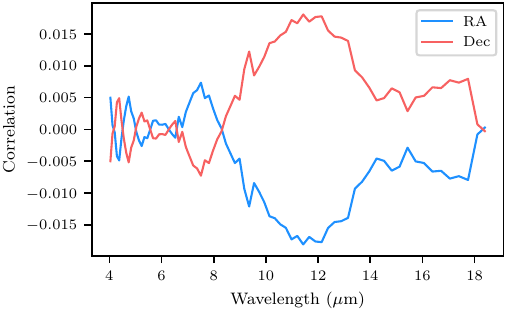}
        \caption{Correlations between the estimated planetary SED and sky coordinates. The correlations do generally depend on the position of the planet and may look differently in other cases.}
        \label{fig:correlation_pos}
    \end{figure}
    In its off-diagonal elements, it shows relatively low correlations in a distinct fringe-like pattern. The magnitude of the correlations depends also on observation parameters and is expected to increase for e.g. lower signal-to-noise ratio cases or cases where not a full array rotation is sampled. This pattern is also observed in the covariance or correlation matrix of the pure planet signal, such as the one displayed in Figure \ref{fig:photon_counts_correlations} (top). This indicates that the covariance of the planet signal propagates through the signal extraction procedure and may also explain the strong oscillation pattern we have observed in the extracted SED of the unwhitened case in Figure \ref{fig:sed_systematic_error} (top). For completeness, we also show the correlations between the SED and the planetary position in Figure \ref{fig:correlation_pos}. Note that the nature of these correlations does generally depend on the planet position.\par
    Previous studies for the retrieval of exoplanet atmospheres in the context of LIFE have assumed that the errors across the wavelength bins of the planetary SED are uncorrelated and independent \citep[e.g.][]{konrad_large_2022, alei_large_2022}. This corresponds to considering only the diagonal terms of the spectral covariance matrix, which has been calculated analytically using \textsc{LIFEsim} \citep{dannert_large_2022}. However, several studies have demonstrated an increase in the quality of retrieved parameters upon consideration of the full spectral covariance \citep[e.g.][]{Greco_measurement_2016, nasedkin_impacts_2023}. In this regard, implementing the numerical method for an estimation of the full spectral covariance matrix, as outlined in Section \ref{sec:data_processing_signal_extraction}, could contribute to a next step in the development of atmospheric retrievals for the LIFE mission in the presence of instrumental errors. Finally, these spectral covariances will also be interesting in the context of multiplanet systems, as we expect also correlations between the spectra of different planets.



%
%

    \section{Summary \& Conclusion}\label{sec:conclusion}
    We have addressed the role of correlated errors in the specification of technical requirements for perturbed space-based nulling interferometers. We have presented a framework that quantitatively connects instrumental perturbations to performance metrics, accounting for the covariance of instrumental errors. Our findings show that data whitening is crucial for improving the sensitivity of detection tests and the quality of the extracted planetary properties.\par
    Previous studies of the performance of such instruments have used assumptions about the independence of their data, which are known to be invalid. With the framework presented here, we can reassess the technical specifications using solid hypothesis testing criteria. Furthermore, our approach allows an estimate of the correlations of the posterior distribution of the astrophysical parameters so that they can be taken into account in atmospheric retrievals. In the context of the LIFE mission, \textsc{LIFEsimMC} will serve as an essential tool to support the derivation of requirements. \textsc{LIFEsimMC} also demonstrates the potential of \textsc{PHRINGE} to integrate into end-to-end simulation pipelines with its state-of-the-art synthetic data generation capabilities and support for the upcoming novel nulling data standard NIFITS.\par
    More research is required to estimate the instrumental error covariance used in the whitening transformation directly from the data, reducing the reliance on observations from calibration stars. This will require efficient means to disentangle the planet from the noise. Further work is also required to account for temporal correlations and non-Gaussian noise in the whitened data to further improve the usability of the test statistics. A first step in this direction could involve integrating the approach of \cite{dannert_consequences_2025} into the framework presented in this work. 

\begin{acknowledgments}
    We thank the anonymous referee for the valuable feedback, which has significantly contributed to the clarity of this paper. This work has been carried out within the framework of the NCCR PlanetS supported by the Swiss National Science Foundation under grants 51NF40\_182901 and 51NF40\_205606. This work has received funding from the Research Foundation -  Flanders (FWO) under the grant number 1234224N. We thank Markus J. Bonse for the support in the development of \textsc{PHRINGE} and numerous helpful discussions that contributed to the outcome of this manuscript.\par
    \textit{Author Contributions:} P.A.H. carried out the analyses made in this work, wrote the manuscript, and, with support of F.A.D. and L.R., developed \textsc{PHRINGE} and \textsc{LFIEsimMC}. F.A.D and R.L. significantly contributed to the development of the presented framework and focus of this manuscript. S.P.Q. and A.M.G. supported this work by providing access to essential resources. All authors commented on the manuscript.\par
    \textit{Software:} We have developed and used \textsc{PHRINGE} \citep{huber_2025_phringe} and \textsc{LIFEsimMC} \citep{huber_2025_lifesimmc}. Both tools rely on several open-source Python packages, among them Astropy \citep{The_Astropy_Collaboration_2022}, PyTorch \citep{paszke2019pytorchimperativestylehighperformance}, SciPy \citep{2020SciPy-NMeth}, SymPy \citep{sympy-10.7717/peerj-cs.103}, SpectRes \citep{carnall2017spectresfastspectralresampling}, and lmfit \citep{matt_newville_2024_12785036}. To support their development, we have also used GitHub Copilot and ChatGPT. Finally, we have used ChatGPT to improve the grammar and style of selected sentences.
\end{acknowledgments}

%





    \appendix

    \section{Astrophysical Noise Sources}\label{sec:appendix_astro_noise}

    This section contains an overview of the astrophysical noise sources used in \textsc{PHRINGE} and \textsc{LIFEsimMC}.

    \subsection{Host Star}
    We model the star as a time-independent uniform disk of radius $r_\star$ with the spectral radiance of a blackbody given by the Planck law. In particular, we do not consider effects such as limb darkening. Assuming the small-angle approximation for the angular radius of the star, $r_\star/d_\star$, its sky brightness is given by
    \begin{equation}
        I_{\star}(\lambda, \boldvec{\alpha})=\Theta\left(\frac{r_\star}{d_\star}-|\boldvec{\alpha}|\right)B(\lambda, T_\star),
    \end{equation}
    where $\Theta$ is the Heaviside step function accounting for the disk shape of the star, $d_\star$ the distance between the observatory and the star and $B$ is the spectral radiance in units of \si{ph.s^{-1}.m^{-2}.\mu m^{-1}.sr^{-2}} as defined by the Planck law:
    \begin{equation}\label{eq:planck}
        B(\lambda, T)= \frac{2hc^2} {\lambda^5}\frac{1}{
                \exp\left(hc/\left(k \lambda T\right)\right) - 1}.
    \end{equation}
    Here, $h$ the Planck constant, $c$ the speed of light, and $k$ the Boltzmann constant.
    
    \subsection{Local Zodiacal Dust}
    The zodiacal dust found mostly near the ecliptic plane in our Solar system acts as a source of background radiation for our measurement. We use the sky position-dependent model by \cite{den_hartog_darwinsim_2005} to model this contamination, which, in an ecliptic coordinate system, gives the sky brightness distribution
    \begin{equation}\label{eq:appendix_local_zodi}
        \begin{aligned}
            I_{\text{LZ}}(\lambda, \boldvec{\alpha})\;\hat{=}&\;I_{\text{LZ}}(\lambda, \lambda_\text{rel}, \beta_\text{lat}) \\=& \tau \left(B(\lambda, T_\text{eff})+AB_\lambda(T_\odot)\left(\frac{r_\odot}{\SI{1.5}{AU}}\right)^2\right)\\
            &\cdot\left(\frac{\pi/\arccos\left(\cos \lambda_\text{rel} \cos\beta_\text{lat}\right)}{\sin^2\beta_\text{lat} + 0.6 \cdot\left(\lambda/(\SI[exponent-product=\ensuremath{\cdot}]{11e-6}{m})\right)^{-0.4}\cos^2\beta_\text{lat}}\right)^{0.5},
        \end{aligned}
    \end{equation}
    where $\lambda_\text{rel}=\lambda_\text{ecl}-\lambda_{\text{ecl}, \odot}$ is the ecliptic longitude relative to the Sun, $\beta_\text{lat}$ is the ecliptic latitude, $\tau=4\cdot 10^{-8}$ is the optical depth toward the ecliptic poles, $T_\text{eff}=\SI{265}{K}$ is the effective temperature of the local zodiacal dust at \SI{1}{AU} from the Sun, $A=0.22$ is the near-IR dust albedo, $T_\odot=\SI{5778}{K}$ is the temperature of the Sun and $r_\odot$ is the radius of the Sun in AU.

    \subsection{Exozodiacal Dust}
    Similar to the zodiacal dust in our own Solar system, exoplanetary systems are expected to contain zodiacal dust. We follow the same approach as \cite{dannert_large_2022} and use the model by \cite{kennedy_exo-zodi_2015} to account for the exozodiacal dust in our simulations. We thus model it as a face-on disk surrounding the host star of the planet, using a power-law distribution for its surface density, $\sigma$, and a radius-dependent temperature profile, $T_\text{EZ}$. This yields the following sky brightness distribution:
    \begin{equation}
        I_{\text{EZ}}(\lambda, \boldvec{\alpha})= \sigma (\boldvec{\alpha})B(\lambda, T_{EZ}(\boldvec{\alpha})).
    \end{equation}
    The unitless surface density is defined as 
    \begin{equation}
        \sigma(\boldvec{\alpha})=7.12 \cdot 10^{-8} z \left(|\boldvec{\alpha}|\;  \frac{d_\star}{\text{m}}\left(\frac{L_\star}{L_\odot}\right)^{0.5}\right)^{-0.34},
    \end{equation}
    where $z$ is the so-called zodi level, a unitless factor indicating a $z$-times higher surface density than the one of the zodiacal dust in our Solar system and $L_\star/L_\odot$ is the luminosity of the star in units of Solar luminosity. The temperature profile is defined as
    \begin{equation}
        T_\text{EZ}(\boldvec{\alpha})=278.3 L_\star^{0.25}|\boldvec{\alpha}|^{-0.5} \SI{}{K}.
    \end{equation}

    \section{Signature for different planetary SEDs}\label{sec:appendix_flux_and_counts}
\begin{figure*}
        \includegraphics[width=\linewidth]{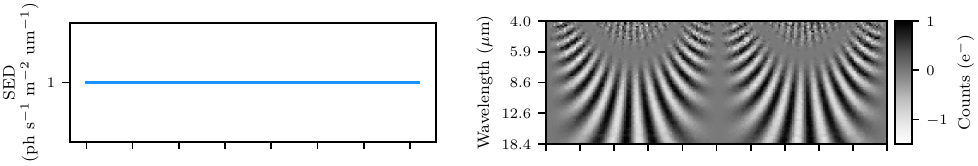}\vspace{0.1cm}
        \includegraphics[width=\linewidth]{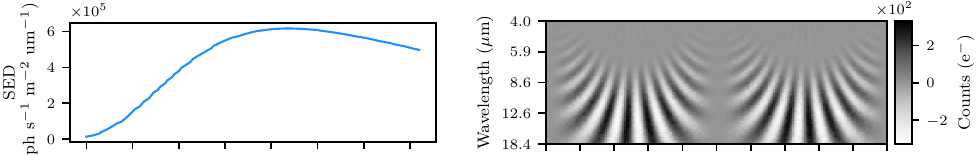}\vspace{0.1cm}
        \includegraphics[width=\linewidth]{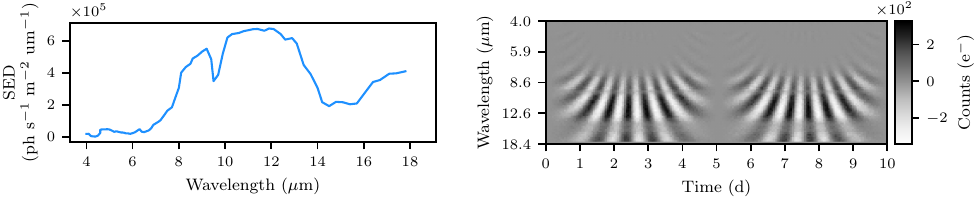}
        \caption{Planetary SED (left) and the corresponding signature in the differential output (right) for an instrument and observation as specified in Table \ref{tab:mission_parameters}. \textit{Top:} Unit SED. The impact of shot noise is visible at the lower wavelengths. \textit{Center:} Blackbody spectrum for $T_\text{p}=\SI{254}{K}$. Compared to the unit SED one can clearly see the reduction in signal at the lower wavelengths. \textit{Bottom:} Earth spectrum. Again, the absorption bands in the atmosphere cause a clear reduction in signal at the corresponding wavelengths. Note that different time step binning is used than in Figure \ref{fig:photon_counts_correlations}, explaining the different values of the color bar.}
        \label{fig:flux_and_counts}
    \end{figure*}

    Figure \ref{fig:flux_and_counts} shows the planetary signature in the differential output of a double Bracewell nuller for three different input spectra: a unit SED (top), a blackbody spectrum for $T_\text{p}=\SI{254}{K}$ (center), and an Earth spectrum (bottom).
    The signature with a unit SED show high photon shot noise at the lower wavelengths. The signature of the blackbody spectrum shows a clear reduction of the signal at lower wavelengths, while the signature of the Earth spectrum shows further reductions at the absorption bands of Earth's atmosphere.

    \section{Complementary Results}
    
    \subsection{Application to M-Type Star Systems}\label{app:subsec_m_stars}
\begin{figure}
        \centering
        \includegraphics[width=\linewidth]{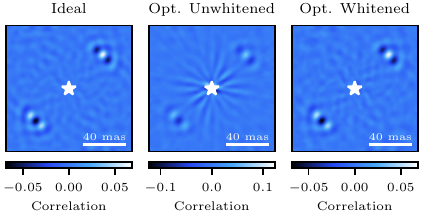}
        \caption{Correlation maps of the Earth-like planet around an M star for the ideal (left), optimistic unwhitened (center) and optimistic whitened (right) cases.}
        \label{fig:image_m_star}
    \end{figure}

    In this study we have focused on a particular reference case of an Earth twin around a Sun-like, i.e. G-type, star at \SI{10}{pc}. However, LIFE is also capable of detecting and characterizing planets orbiting M-type stars \citep{quanz_large_2022}. We thus repeat a part of our analysis for an M0V star at \SI{10}{pc} with $M_\star=\SI{0.57}{M_\odot}$, $R_\star=\SI{0.588}{M_\odot}$, $T_\star=\SI{3850}{K}$ \citep{mamajek_modern_2022} with an Earth twin with a semi-major axis of $a_\text{p}=\SI{0.377}{AU}$. Figure \ref{fig:image_m_star} shows the correlation maps for such an observation, given an ideal (left) and an optimistic amount of instrumental noise, as specified in Table \ref{tab:scenarios}, without (middle) and with (right) the application of the ZCA whitening transformation. 
    As expected, the smaller angular size and lower flux of the star lead to better performance than for the G-type star under otherwise identical conditions.

    \subsection{Additional Spectra}
\begin{figure*}
        \includegraphics[width=\linewidth]{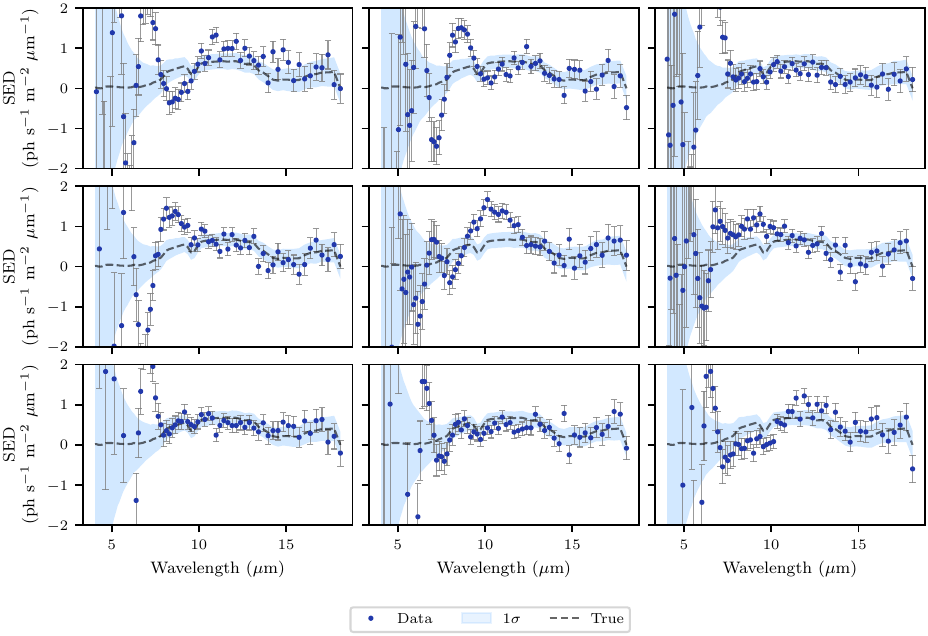}
        \caption{Additional examples of extracted planetary SEDs with corresponding uncertainties for the optimistic unwhitened case, all showing clear systematic errors below $\sim \SI{10}{\mu m}$. Variations among the different panels stem from the random instantiations of the instrumental perturbations.}
        \label{fig:additional_spectra}
    \end{figure*}

    Figure \ref{fig:additional_spectra} shows nine additional spectra corresponding to the extracted planetary SEDs for the optimistic unwhitened case, supplementing Figure \ref{fig:sed_systematic_error}.
    

\bibliography{bib}{}
\bibliographystyle{aasjournal}



\end{document}